\documentclass[fleqn,10pt]{wlscirep}

\usepackage{booktabs} 

\usepackage[ruled]{algorithm2e} 

\usepackage{graphicx}
\usepackage{todonotes}
\usepackage{textcomp}
\usepackage{caption}
\usepackage{subcaption}
\usepackage{todonotes}
\usepackage{color}
\usepackage{float}

\begin{document}

\title{Web Video in Numbers\\ An Analysis of Web-Video Metadata}  
\author{Luca Rossetto}
\author{Heiko Schuldt}

\affil{University of Basel, Department of Mathematics and Computer Science\\ Spiegelgasse 1, 4051 Basel, Switzerland\\
luca.rossetto@unibas.ch \quad ORCID: \href{http://orcid.org/0000-0002-5389-9465}{0000-0002-5389-9465}\\
heiko.schuldt@unibas.ch}

\begin{abstract}
Web video is often used as a source of data in various fields of study. While specialized subsets of web video, mainly earmarked for dedicated purposes, are often analyzed in detail, there is little information available about the properties of web video as a whole. In this paper we present insights gained from the analysis of the metadata associated with more than 120 million videos harvested from two popular web video platforms, vimeo and YouTube, in 2016 and compare their properties with the ones found in commonly used video collections. This comparison has revealed that existing collections do not (or no longer) properly reflect the properties of web video ``in the wild''.
\end{abstract}

\maketitle

\section{Introduction}
Over recent years, video has become a significant portion of the overall data which populates the web. Web video has therefore become relevant in multiple areas of research, especially in the context of multimedia and data management. For these areas, it is important to have datasets people can refer to so as to ensure comprehensible and reproducible results. 
Existing video collections suffer from the fact that they are earmarked for certain applications, thus do not reflect a very broad spectrum of content and/or they are subject to copyright constraints so that they cannot be freely used and redistributed without legal restrictions. 
As a consequence, there is very little information about the properties of web video as a whole which makes it very hard to assess the representativeness of such collections.

In this paper, we explore the properties of metadata of web video in the wild as found on two of the largest web video platforms and compare them to the properties of large, commonly used video collections. To do this, we collected the metadata of 20 million videos from vimeo\footnote{\url{https://vimeo.com}} and of 100 million YouTube\footnote{\url{https://youtube.com}} videos. This collection process was performed between April and July of 2016 and has returned videos created between 2005 and 2016.

The contributions of this paper are twofold: first, we present the analysis results of the vimeo and YouTube content we have harvested, giving detailed insights into the properties of web video in the wild as of mid 2016. Second, we show how these general purpose collections differ from existing collections which thus do not properly reflect the properties of the video content that can be found on the web.

The rest of this paper is structured as follows: Section~\ref{sec:related} provides an overview of commonly used multimedia collections and their areas of application. In Section~\ref{sec:collection} we elaborate on the methods used to collect the metadata used in the analysis, which is presented in Sections~\ref{sec:analysis} and \ref{sec:more_analysis}. Section~\ref{sec:discussion} discusses the analysis results and Section~\ref{sec:dataset} details the structure of the dataset and provides information on how to obtain it. Section~\ref{sec:conclusion} concludes.

\pagebreak

\section{Related Work}
\label{sec:related}

Web video gained importance in various areas within the sciences and humanities in recent years. In computer science, it is mostly used in the context of retrieval~\cite{da2016near} and machine learning~\cite{cao2016web,li2016online}. To facilitate research in these areas, multiple video collections have been compiled over the years. The majority of these datasets was created with a specific problem in mind and they are therefore comprised of a comparatively small number of videos with a narrow range of content. Such collections are usually used for the evaluation of concept recognition tasks \cite{das2013thousand} or action recognition tasks \cite{ryoo2013first}. Other, more general collections range in size from a few tens \cite{geisler2000open,OSVC1} to many thousands of hours of video content \cite{over2009creating,song2011multiple,thomee2015yfcc100m}. The more popular of these include CC\_WEB\_VIDEO~\cite{wu2009real}, MCG-WEBV~\cite{cao2009mcg} and IACC~\cite{over2009creating}. What these collections have in common is that they all source their videos from the web, but all of them have their own criteria for inclusion of videos. This introduces biases which prevents these collections from being representatives for the state of web video as a whole.

Some of these collections are limited in their applicability as they contain content whose licenses do not explicitly allow a free and unconstrained use \cite{marszalek09}. Such licensing issues can lead to situations where collections which were purpose-built for certain applications and are used by several researchers cannot easily be shared with the research community at large without the corresponding legal paper work \cite{cobarzaninteractive}. Other collections which are especially built to avoid such issues yet suffer from other limitations. In \cite{over2009creating}, the authors use the Internet Archive\footnote{\url{https://archive.org/details/movies}} as a source for video. While all these videos are guaranteed to be freely usable and re-distributable, they are not representative of modern web video content found in the wild because of a certain lack of diversity in their sources. Other collections, such as~\cite{thomee2015yfcc100m}, avoid this problem by collecting creative commons\footnote{\url{https://creativecommons.org}} licensed images and videos from flickr\footnote{\url{https://www.flickr.com}} which increases the diversity. Due to the primary focus of flicker on images rather than videos, the videos found on this platform still differ from what can currently be regarded as general web video.

Mainly in the context of data mining, the metadata of web video has also been studied. Such works were aiming at predicting certain properties of videos from specific genres~\cite{algur2016web} or at trying to detect specific behaviors such as privacy invasion~\cite{aggarwal2014mining}. Web video metadata has also been used to study video popularity distributions across different video platform as well as copyright infringement~\cite{imc2007cha}. Again, such approaches are usually based on metadata which was collected especially for this purpose and thus does not provide a representative cross section of web video in general.

Other strategies to analyze the video metadata is to observe the traffic from a web video site like YouTube at the edge of the network such as internet gateways of university campuses~\cite{gill2007youtube,zink2009characteristics}. Such approaches can provide great insights into the video viewing behavior of the users using this gateway. The downside is, however, that it provides no information about videos not watched by this particular user group which means that it is also limited in its generality.

\pagebreak

\section{Collection Process}
\label{sec:collection}
This section describes the setup used to collect the video metadata from vimeo and YouTube as well as the strategies that have been employed.

\subsection{Setup}
\begin{figure}[t]
	\centering
	\includegraphics[width=0.8\textwidth]{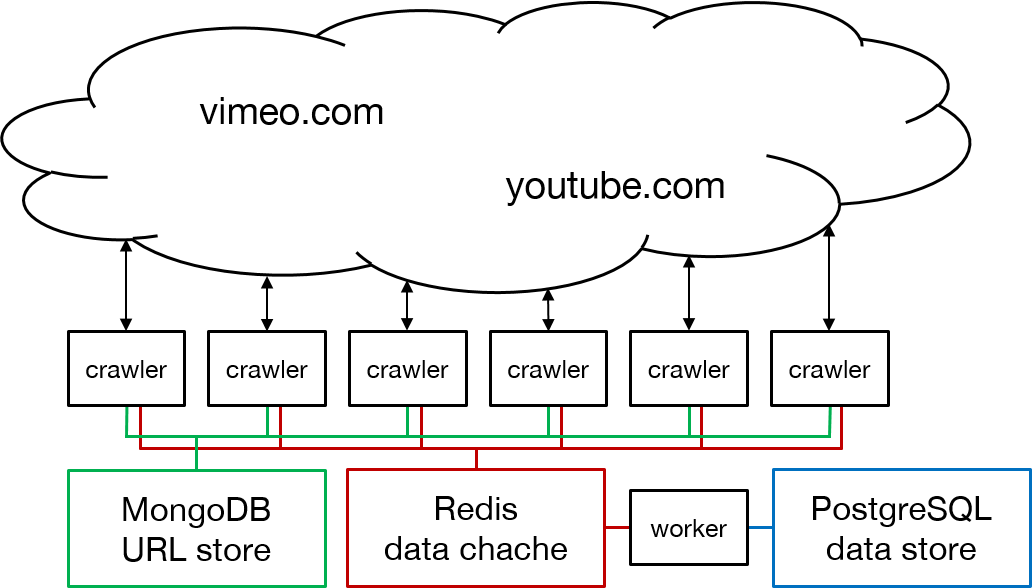}
	\caption{Architecture of the crawling setup}
	\label{fig:collection-process}
\end{figure}

To collect the data from the two platforms, vimeo and YouTube, we implemented a custom distributed crawler. This crawler is implemented in Java and uses jsoup\footnote{\url{https://jsoup.org}} as an HTTP client and HTML parser. Many crawler instances can work in parallel which distributes system load while also reducing the number of requests originating per IP address. The crawler instances are synchronized via a URL-queue stored in a mongodb\footnote{\url{https://www.mongodb.com}} database. A crawling instance selects a random URL from the first 10'000 unvisited URLs, analyses the page and adds the not yet seen URLs to the queue. The actual data collected by the crawlers are added to a temporary distributed queue\footnote{\url{http://redisson.org}} stored in a redis\footnote{\url{http://redis.io}} database from where a dedicated worker takes the data and stores them persistently in a postgres\footnote{\url{https://www.postgresql.org}} database. The architecture used for the collection process is illustrated in Figure~\ref{fig:collection-process}.

\subsection{Crawling Strategy}
Since the two considered sites have very different ways in which they present videos in the larger context of other hosted videos, two different crawling strategies were necessary.

\subsubsection{vimeo}
A page containing a video on vimeo primarily focuses on this particular video and its creator. The page therefore mostly --if not exclusively-- displays other videos by the same creator. The only reliable way we found to get from one video to related videos from different creators is via the tags associated with each video. For each tag, vimeo links to a list of all videos tagged correspondingly which is segmented in pages of up to 10 videos, ordered by upload date. The crawler therefore examines  all the tags of a video and puts the first page of every tag list into the URL-queue. For every tag page, it adds all listed videos as well as the `next' tag page to the URL-queue. Since videos are sorted by upload date showing the newest first rather than oldest first, this strategy causes the crawler to miss videos since the content of a previously visited page will change with every newly added video, pushing previously visited videos back onto as of yet unvisited tag pages. The crawler was seeded with the oldest video on vimeo\footnote{\url{https://vimeo.com/2}}.

\subsubsection{YouTube} 
Since every video page on YouTube contains a list of videos which YouTube considers to be related, it is easy for the crawler to collect a large number of videos from different creators. For every video page visited, the crawler adds the link to the creator's channel as well as a fraction of all related videos to the URL-queue. This fraction was initially set to $1$ and gradually lowered over time until the number of unvisited pages remained roughly stable when using a value of $\frac{1}{4}$. By the time the sum of visited and unvisited entries in the URL-queue exceeded the target of 100 million entries, addition to the queue was stopped. In order to minimize biases, crawling was only performed from IP addresses not associated with any YouTube account. Since there are many ways in which channel pages can list videos of a particular creator, they were not used as a source of videos to visit but only analysed for information about the creator as well as related channels. The crawler was seeded with the first video on YouTube\footnote{\url{https://www.youtube.com/watch?v=jNQXAC9IVRw}}.

\subsection{Limitations}
Due to the architecture of the two considered platforms, certain metadata could not be collected. Especially vimeo makes heavy use of asynchronous loading of data when composing a page for a video and does not consistently display information such as number of views or likes. We have chosen to omit these data from our collection process in order to maintain consistency within the dataset across platforms and also to reduce the number of requests necessary per video.

\section{Analysis and Comparison of Datasets}
\label{sec:analysis}
In this section, we compare certain statistical properties of four commonly used video collections with statistics obtained from the dataset described in Section~\ref{sec:collection}. The collections analyzed in this section are the TrecVid Internet Archive Creative Commons collection (IACC.3~\cite{over2009creating}) which is composed of videos taken from the internet archive, the WEB Video collection of the Multimedia Computing Group at the Chinese Academy of 
Sciences (MCG-WEBV~\cite{cao2009mcg}), containing the monthly most popular videos from YouTube from 2009, the Yahoo Flickr Creative Commons 100 Million dataset (YFCC100M~\cite{thomee2015yfcc100m}) which contains free videos taken from flickr and the Sports-1M Dataset~\cite{KarpathyCVPR14}, a collection of sports related videos from YouTube.

\begin{figure}[h]
	\centering
	\includegraphics[width=\textwidth]{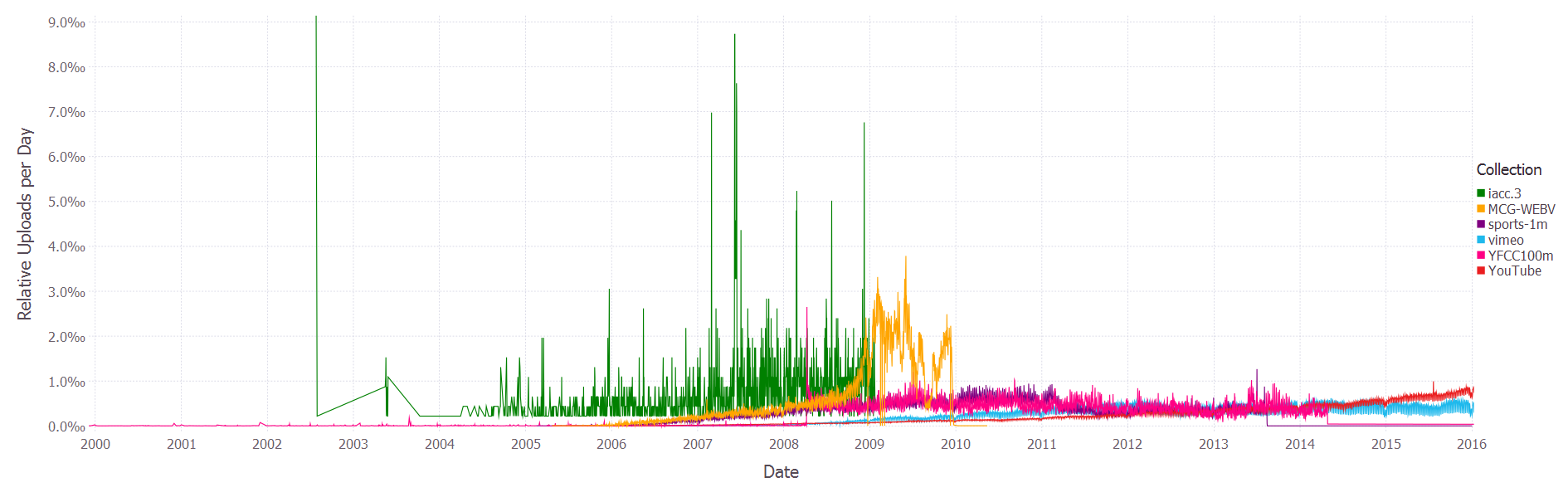}
	\caption{Relative daily video uploads}
	\label{fig:daily}
\end{figure}

\pagebreak

\begin{figure}[h]
	\centering
	\includegraphics[width=\textwidth]{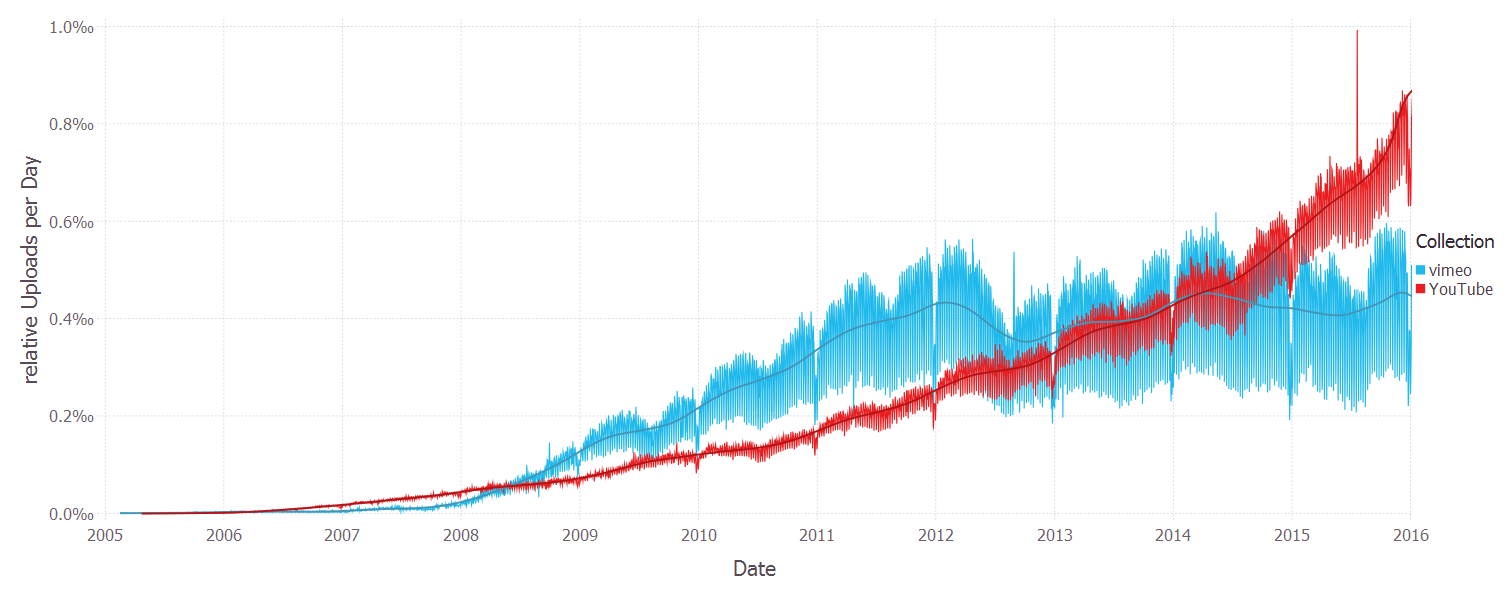}
	\caption{Relative daily video uploads for vimeo and YouTube}
	\label{fig:daily_vy}
\end{figure}

\subsection{Upload date}
To estimate the age distribution of the collections in question, we analyze the reported ages of the contained videos. Since only the YFCC100M meta-data provides the creation date of the video, we use the upload-date for the other collections. Upload dates which pre-date the creation of the platform from which they are collected as well as creation dates before the year 2000 are discarded. 
Figure~\ref{fig:daily} shows the creation/upload rate over time normalized by collection. It can be seen that videos from the IACC.3 are mostly created between 2002 and 2009 and videos from MCG-WEBV are predominantly from 2009. This is not surprising since the MCG-WEBV is comprised of the most popular videos of every month of the year 2009. The oldest videos are from the YFCC100M which is due to the fact that we here consider dates as valid which are older than the platform the videos are hosted on since flickr, being predominantly a photo-sharing platform, differentiates between creation date and upload date. For both the Sports-1M Dataset and the YFCC100M we can see a sharp drop of in age due to the date of the collection process in the years 2013 and 2014 respectively. When compared with the data we obtained from YouTube and Vimeo, we can see that in all of the collections, old videos, in this context meaning from before 2010, are over-proportionally represented, especially in MCG-WEBV and IACC.3. Due to the scaling in Figure~\ref{fig:daily}, some fidelity is lost for the data we collected which span a larger range in time, which is why we show the same data for only the collected metadata in Figure~\ref{fig:daily_vy}. Here it can be seen that while the upload rate for vimeo is relatively stable since 2012, the number of uploads to YouTube is still increasing. The high fluctuations in these graphs are due to differences in the amount of videos uploaded on each weekday.

\begin{figure}[t]
	\centering
	\includegraphics[width=\textwidth]{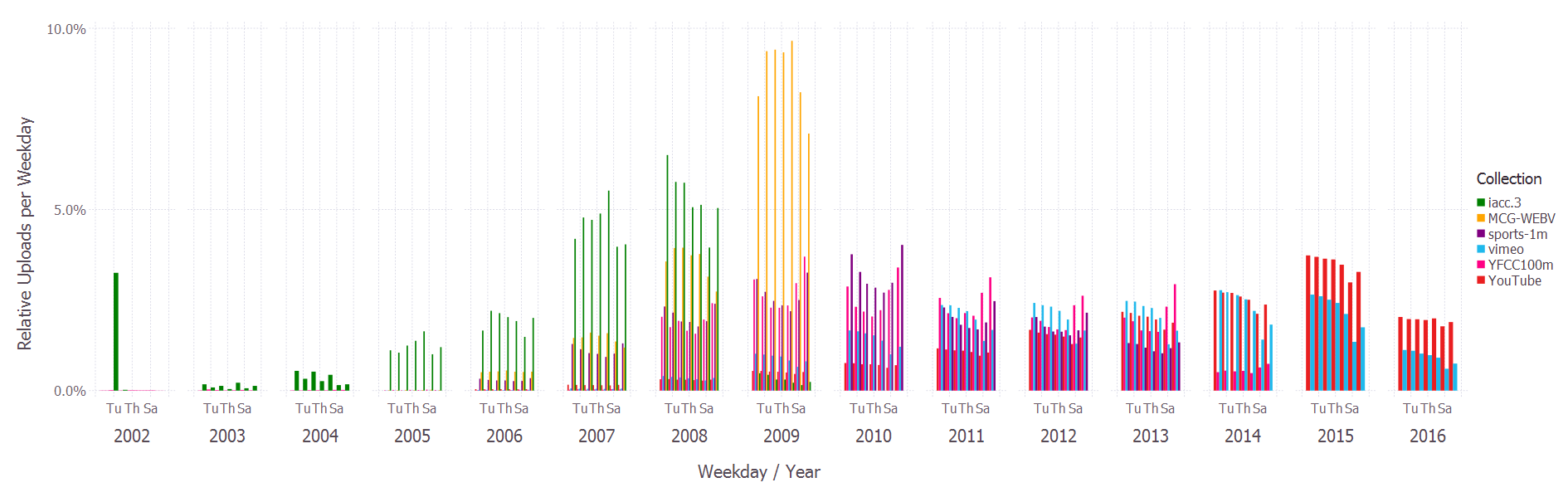}
	\caption{Relative video upload rate per weekday. (Note that the data for 2016 is incomplete since the data collection was performed in Q2 and Q3 of 2016.)}
	\label{fig:weekday}
\end{figure}

\begin{figure}
	\centering
	\includegraphics[width=\textwidth]{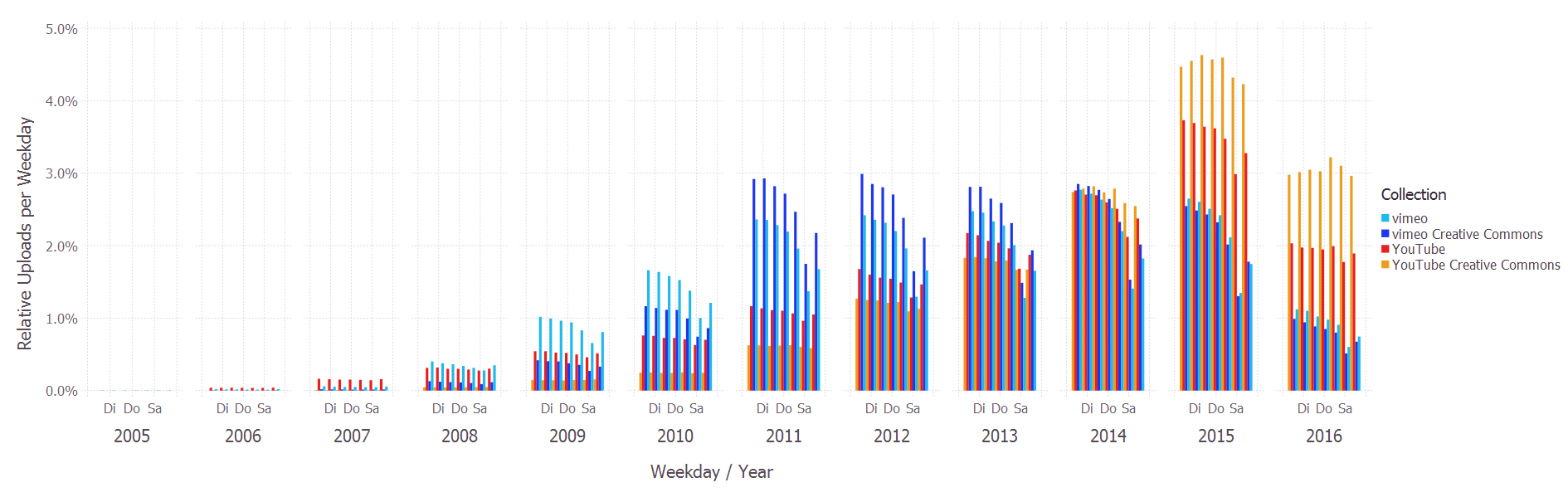}
	\caption{Relative video upload rate per weekday for the collected metadata.}
	\label{fig:cc_weekday}
\end{figure}

\begin{figure}
	\centering
	\includegraphics[width=\textwidth]{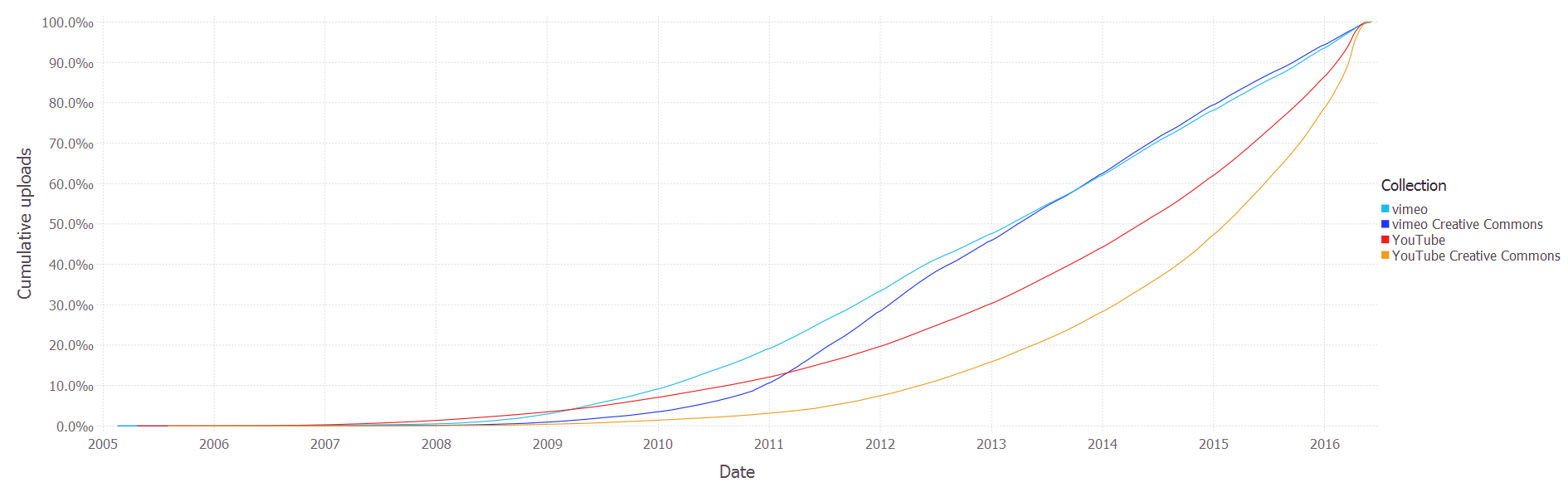}
	\caption{Cumulative video uploads over time}
	\label{fig:cc_upload_cumsum}
\end{figure}

In Figure~\ref{fig:weekday} we look at the distribution of video creations per weekday and year. This data is again normalized by collection but not by year. For the data from vimeo and YouTube a pattern can be observed where most video uploads occur on Monday. From there, the upload rate decreases continuously until it reaches its minimum on Saturday before increasing again. A similar but not as prominent pattern can be observed for the Sports-1M Dataset which is not surprising since its videos are collected from YouTube. It is however interesting that the MCG-WEBV shows no such pattern even though its videos are sourced from YouTube as well. In contrast, it appears that for the YFCC100M, dis-proportionally many videos were created on the weekend. The IACC.3 does not show any discernible pattern with respect to uploads per weekday. To see if these patterns have anything to do with the license under which the videos were published, we show the same analysis for only our collected metadata in Figure~\ref{fig:cc_weekday} and add the data from only the videos released with a creative commons license. For vimeo, the pattern appears independently of whether the video is cc-licensed or not but for YouTube, the peek in upload activity shifts from the beginning towards the middle of the week when only free videos are considered while the trough on weekends disappears. It can also be seen that cc-videos tend to be newer, especially on YouTube, which is more effectively illustrated in Figure~\ref{fig:cc_upload_cumsum}.

\clearpage

\subsection{Duration}
\begin{figure}[h]
	\centering
	\includegraphics[width=\textwidth]{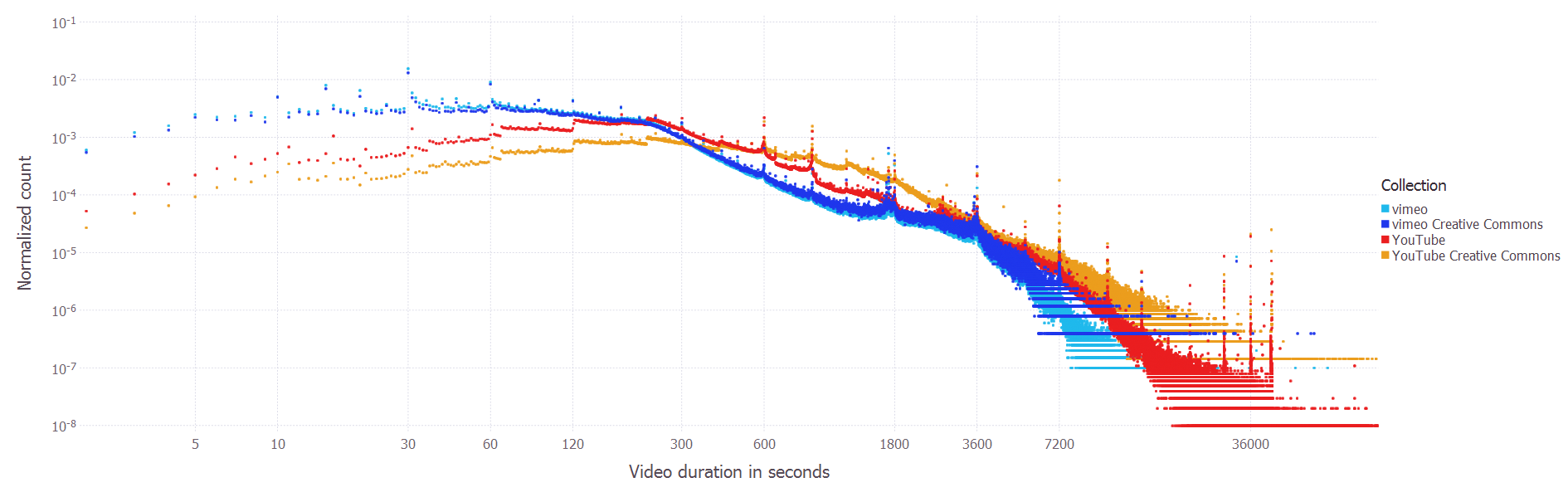}
	\caption{Distribution of video duration}
	\label{fig:cc_duration}
\end{figure}

\begin{figure}[h]
	\centering
	\includegraphics[width=\textwidth]{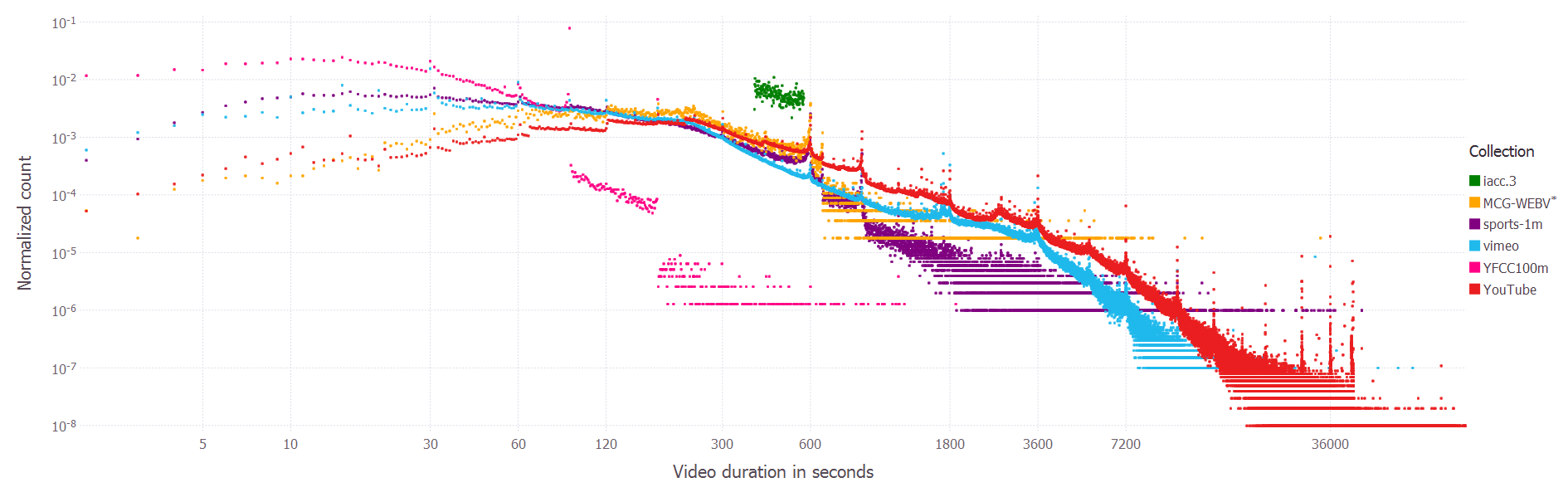}
	\caption{Distribution of video duration}
	\label{fig:duration}
\end{figure}

\begin{figure}[h]
	\centering
	\includegraphics[width=\textwidth]{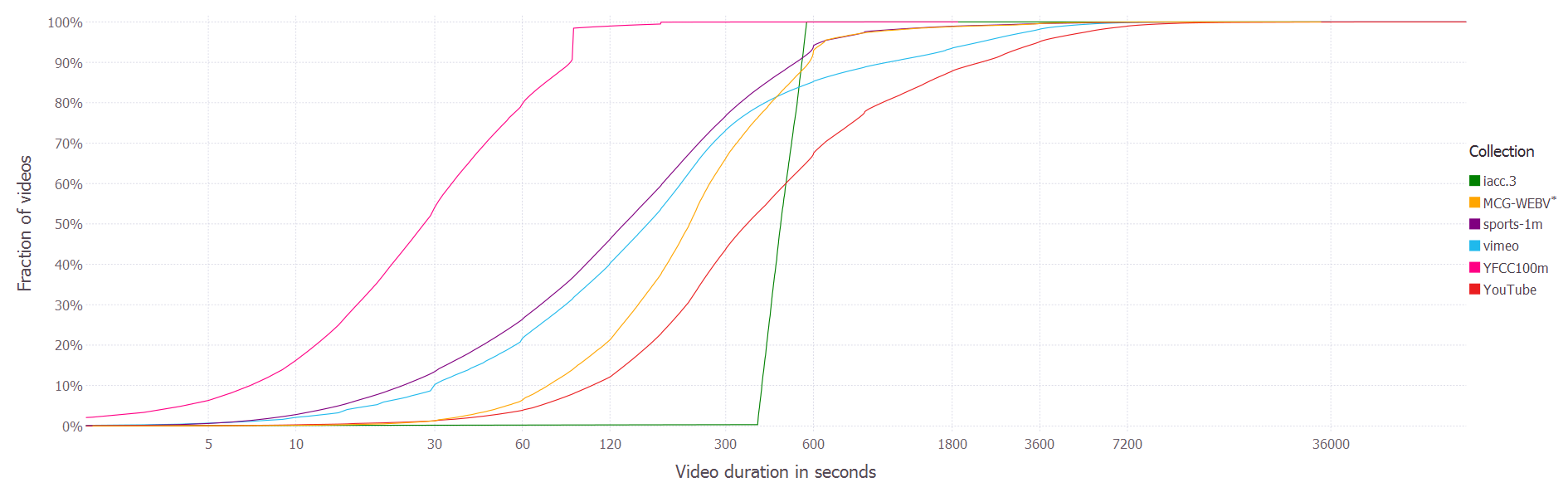}
	\caption{Cumulative distribution of video duration}
	\label{fig:duration_cumsum}
\end{figure}

\begin{figure}[h]
	\centering
	\includegraphics[width=\textwidth]{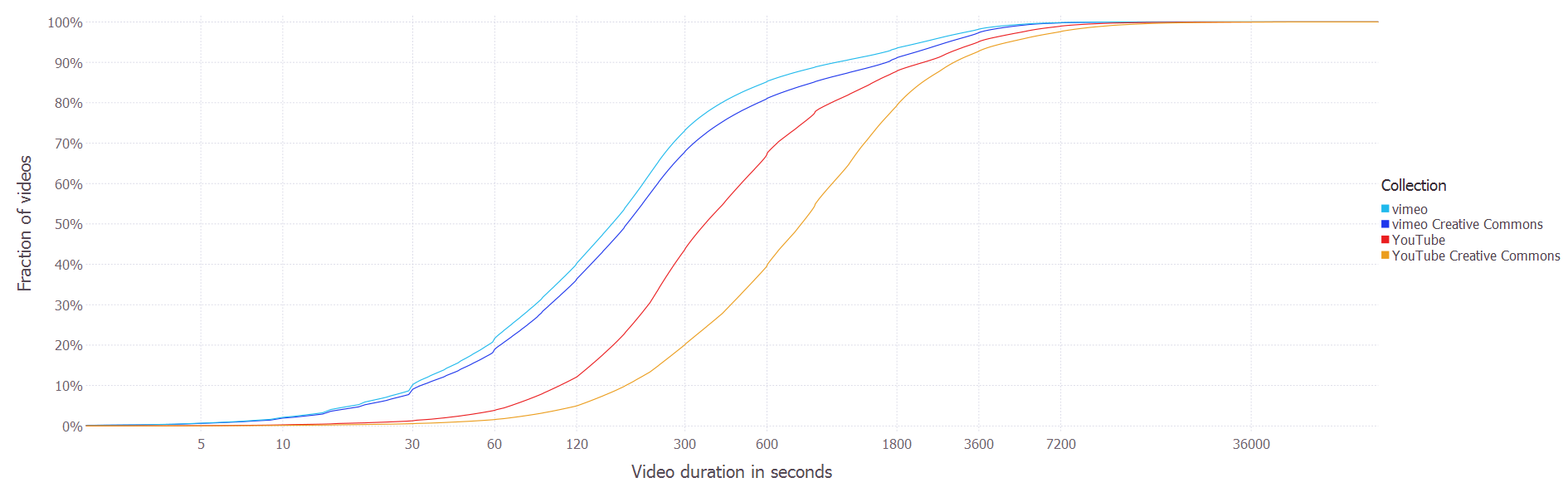}
	\caption{Cumulative distribution of video duration}
	\label{fig:cc_duration_cumsum}
\end{figure}

\begin{figure}[h]
	\centering
	\includegraphics[width=\textwidth]{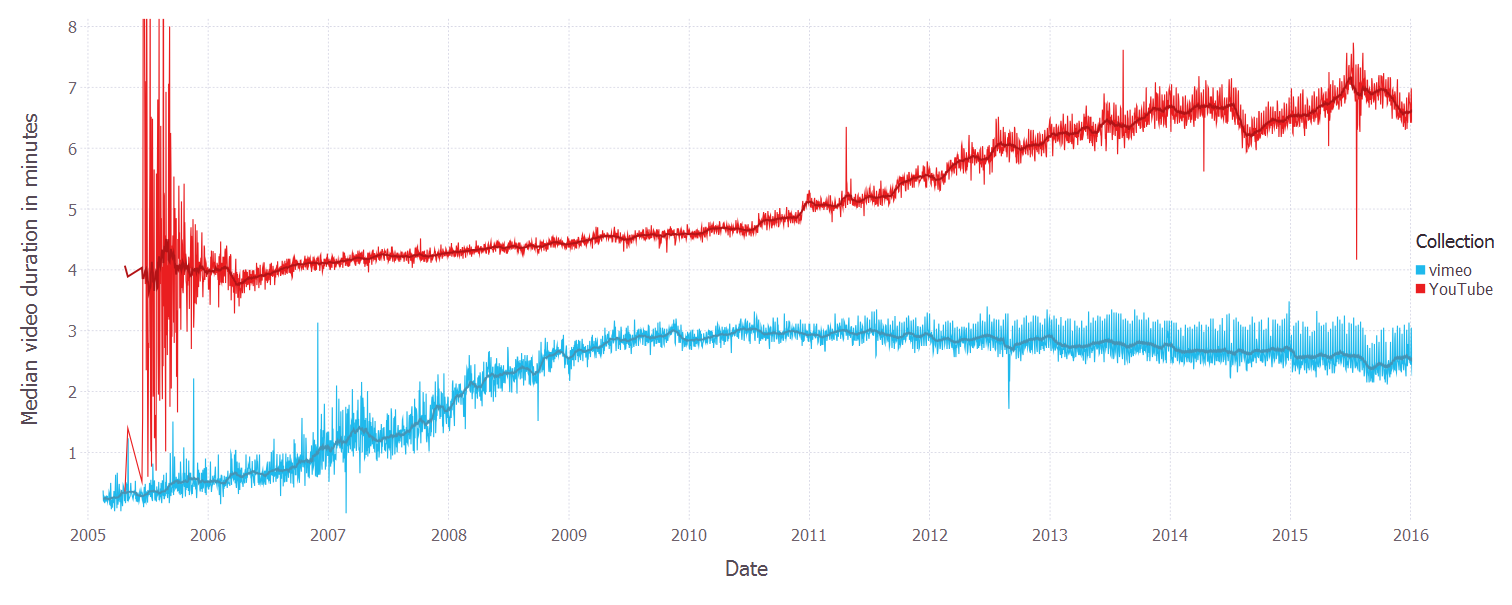}
	\caption{Daily median video duration over time and 30 day moving average}
	\label{fig:duration_vs_time}
\end{figure}

We show the distribution of video duration in Figure~\ref{fig:duration}. Since no duration information was provided in the metadata for MCG-WEBV, IACC.3 and YFCC100M, we had to use external data. For MCG-WEBV the sampling consists of the data from the videos contained both in the collection and our YouTube dataset\footnote{Since this data is incomplete, the collection is denoted with a `*' in the figure. This notation will be used in the rest of the paper from here on.} and for the YFCC100M we used the data provided by~\cite{choi2016analysis}. For the IACC.3, we downloaded the entire collection and use the duration information reported by ffmpeg. It can be seen that most collections show a wide range in duration with the notable exception of IACC.3 and YFCC100M. It can also be seen that while the Sports-1M Dataset does span almost the same range in duration as the data from vimeo and YouTube, the majority of videos are considerably shorter.
This is more apparent in Figure~\ref{fig:duration_cumsum} which shows the cumulative distribution of the duration. It can be seen here that, for example, roughly 80\% of videos of the YFCC100M are at or below 60 seconds in length while vimeo has only about 20\% of videos with one minute or less in length. It can also be seen that all the analyzed collections have a higher percentage of videos at or below 10 minutes in length as both the reference from vimeo and YouTube, begging the conclusion that all these collections are mostly comprised of videos which tend to be shorter than what is found in the wild. As with the previous analyses, we show the same data for our collection and its cc-parts in Figures~\ref{fig:cc_duration} and \ref{fig:cc_duration_cumsum}. It can be seen that for both platforms, free videos are longer, but while the difference on vimeo is rather small, it is more prominent on YouTube. When looking at video duration over time, as depicted in Figure~\ref{fig:duration_vs_time}, we can see that while videos on YouTube have gotten longer in recent years, those on vimeo have slightly declined in length in the median.

\clearpage

\subsection{Resolution}
\label{sec:resolution}
\begin{figure}[h]
	\centering
	\includegraphics[width=\textwidth]{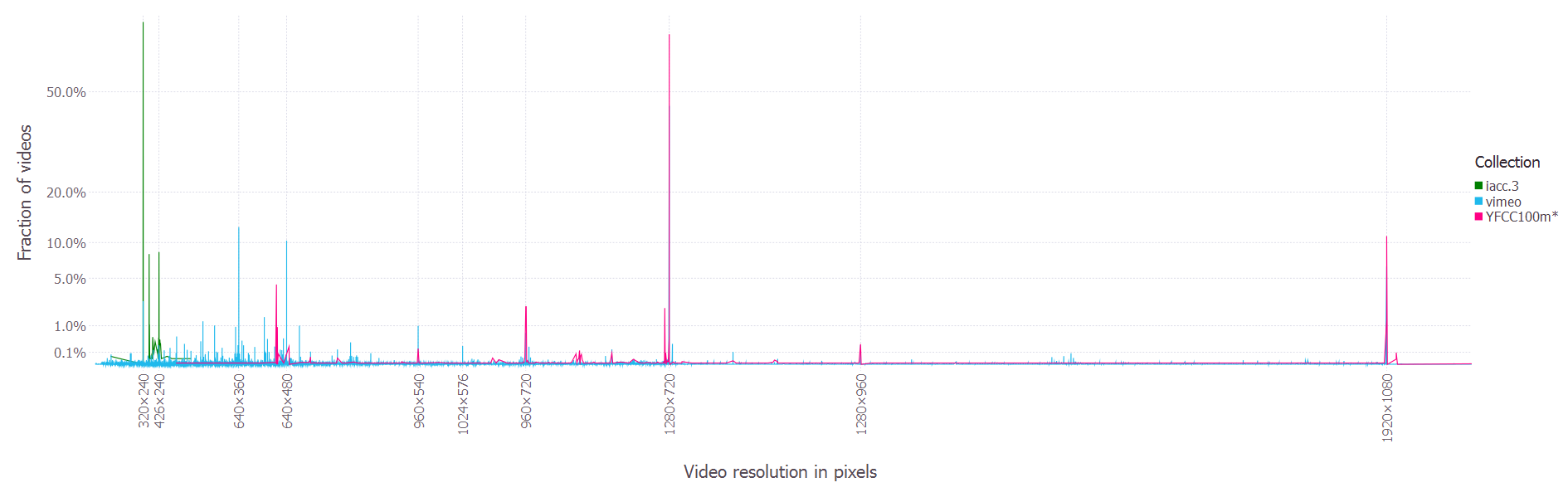}
	\caption{Distribution of video resolution}
	\label{fig:resolution}
\end{figure}

\begin{figure}[h]
	\centering
	\includegraphics[width=\textwidth]{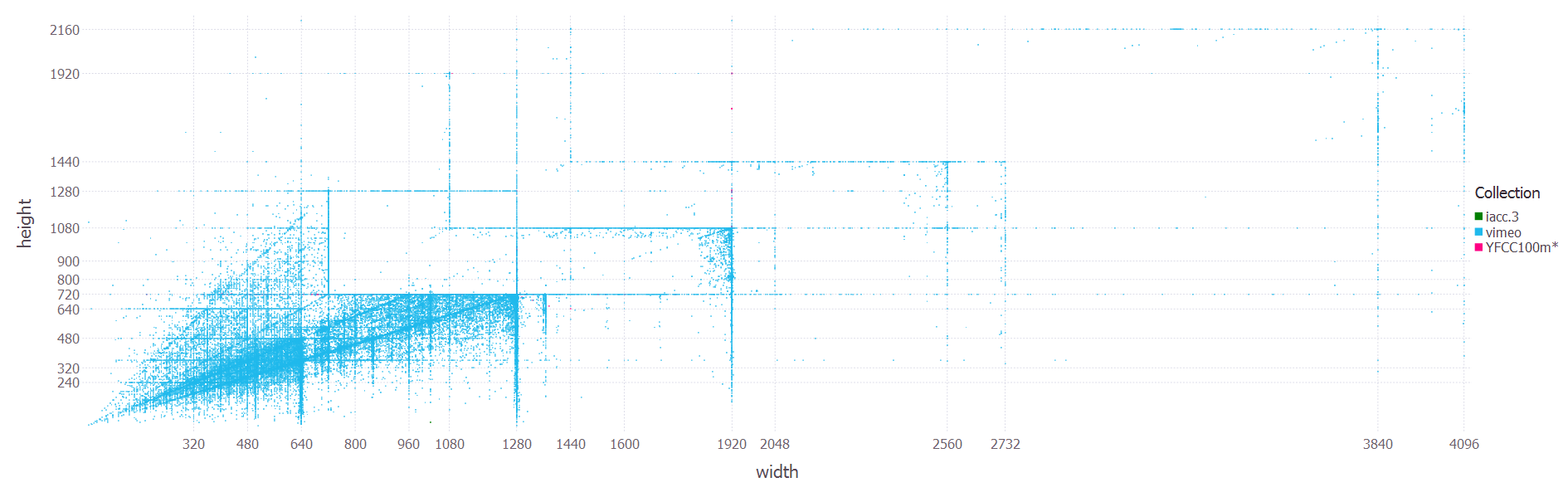}
	\caption{Occurrence of video resolutions}
	\label{fig:vimeo_resolution_scatter}
\end{figure}

\begin{figure}[h]
	\centering
	\includegraphics[width=\textwidth]{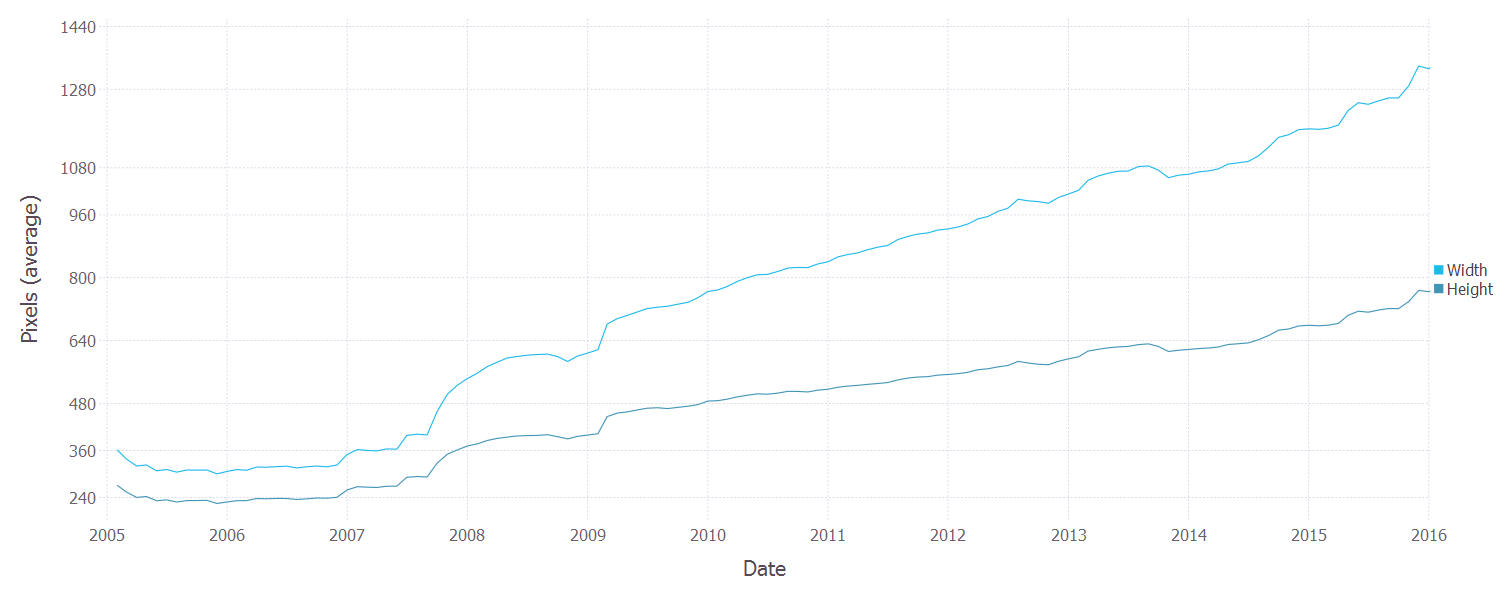}
	\caption{Daily average video resolution on vimeo over time}
	\label{fig:vimeo_res_time}
\end{figure}

\begin{figure}[h]
	\centering
	\includegraphics[width=\textwidth]{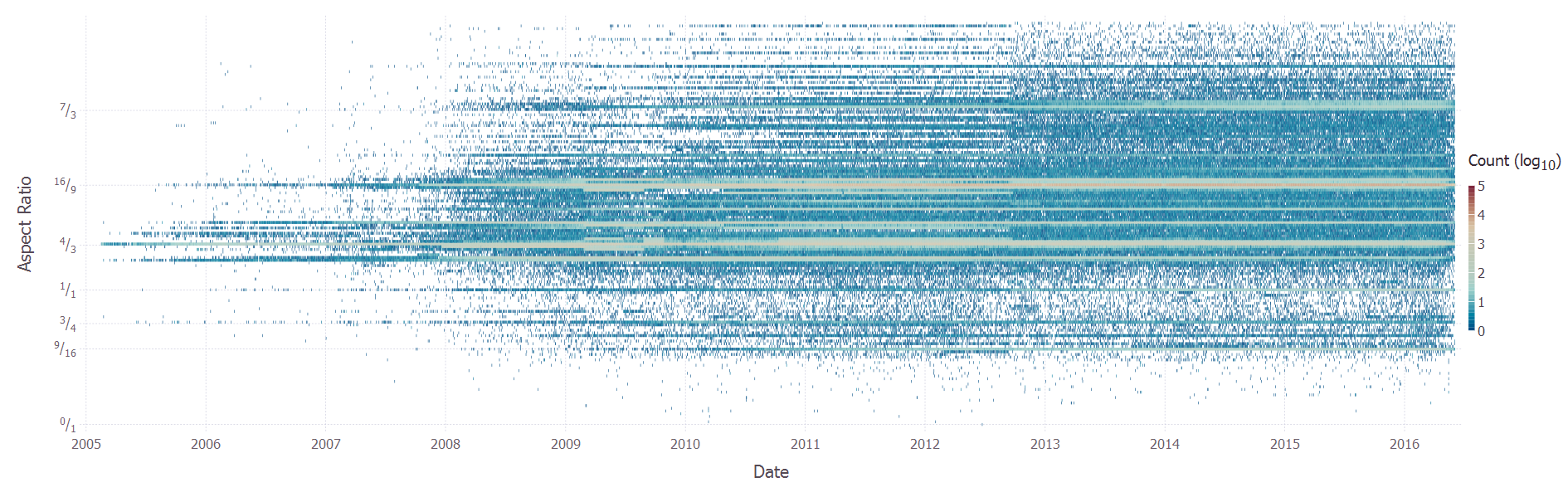}
	\caption{Daily distribution of aspect ratio over time}
	\label{fig:vimeo_aspect_time}
\end{figure}

We compare the distribution of video resolutions by showing the histogram of frame sizes which are obtained by multiplying a frames width by its height. Because most videos have an aspect ratio of (close to) 4:3 or 16:9, there is little error in this representation which would be introduced by two or more commonly occurring but different resolutions having an identical number of pixels per frame. Since YouTube does not support arbitrary video resolution but rather re-scales any uploaded video to a set of pre-defined ones, we will exclude all YouTube based collections from this analysis. Figure~\ref{fig:resolution} shows this distribution for the IACC.3, the YFCC100M and the data from vimeo. Since no resolution information was available in the metadata provided for the YFCC100M, we downloaded the first 80k videos which were still available via the download link listed in the provided metadata and determined their resolutions using ffmpeg. Since the order of the videos within the dataset is randomized, we expect that the distribution obtained in this way approximates the distribution of the full, original collection. It can be seen that the IACC.3 is comprised of videos with both the lowest resolutions and the smallest spread in resolution. While the spread for the YFCC100M is smaller than what we observe for vimeo (which might be in part due to our sub-sampling of this collection), it is still considerably larger than for IACC.3. While both the YFCC100M and the vimeo data have their respective highest peaks at what is commonly referred to as `HD resolution' (1280$\times$720 pixels), YFCC100M has a higher percentage of videos above this resolution. Since the very large number of different resolutions found on vimeo can not accurately be represented in Figure~\ref{fig:resolution}, we show it as a scatter plot in Figure~\ref{fig:vimeo_resolution_scatter}. This figure also shows the resolutions found in the IACC.3 and the YFCC100M which are however less diverse. We also show the daily average video width and height for vimeo in Figure~\ref{fig:vimeo_res_time} which shows that over time, the resolution not only increases but the aspect ratio increases as well resulting in videos which become wider with respect to their height. The distribution of aspect ratio over time is detailed in Figure~\ref{fig:vimeo_aspect_time} which shows the log-amount of the daily occurrence of aspect ratios. It can be seen that not only the wider aspect ratios such as 16:9 and 21:9 increase over time but that videos in portrait mode with ratios of 9:16, which are presumably recorded using smart phones, show a similar growth.

\clearpage

\subsection{Tags}
\begin{figure}[H]
	\centering
	\includegraphics[width=\textwidth]{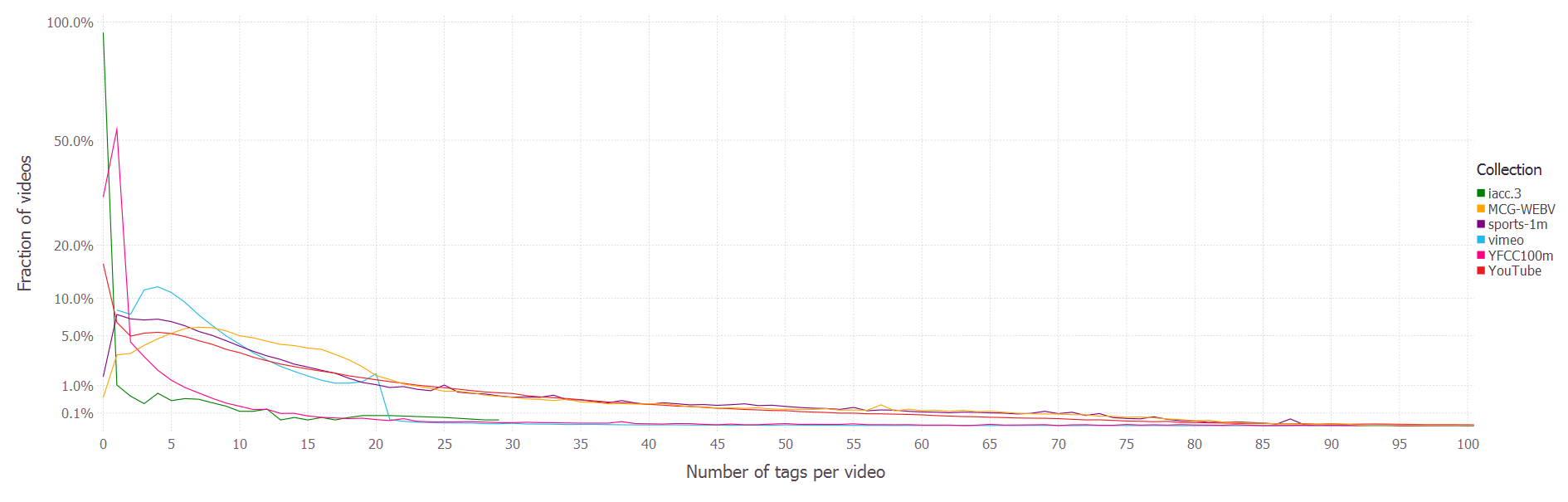}
	\caption{Distribution of tags per video}
	\label{fig:tags_per_video}
\end{figure}

\begin{figure}[H]
	\centering
	\begin{tabular}{cc}
		\includegraphics[width=0.45\textwidth]{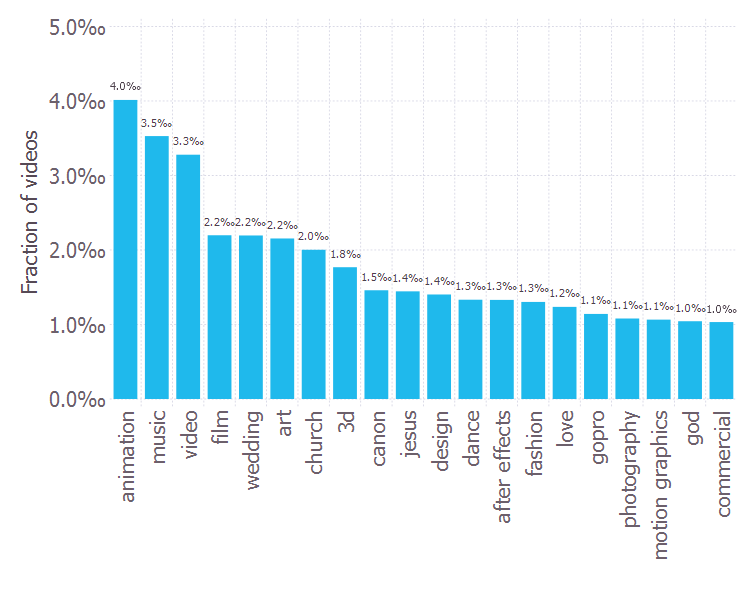} &
		\includegraphics[width=0.45\textwidth]{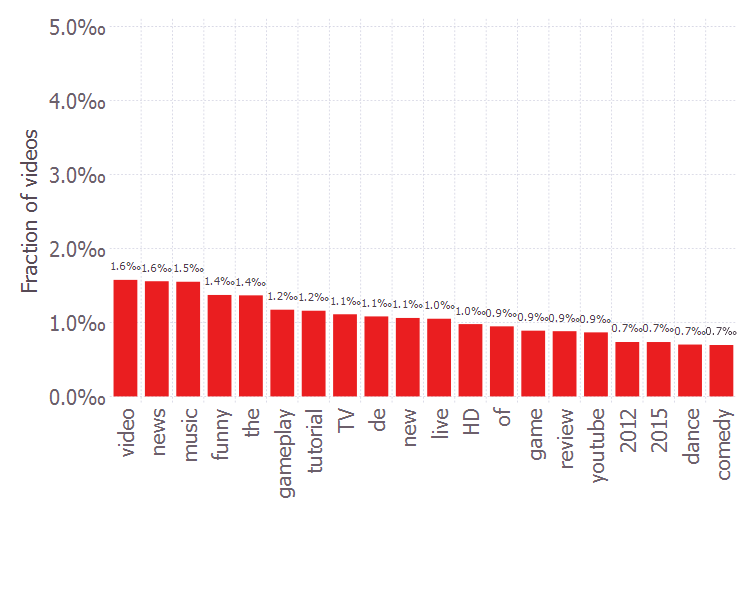} \\
		\includegraphics[width=0.45\textwidth]{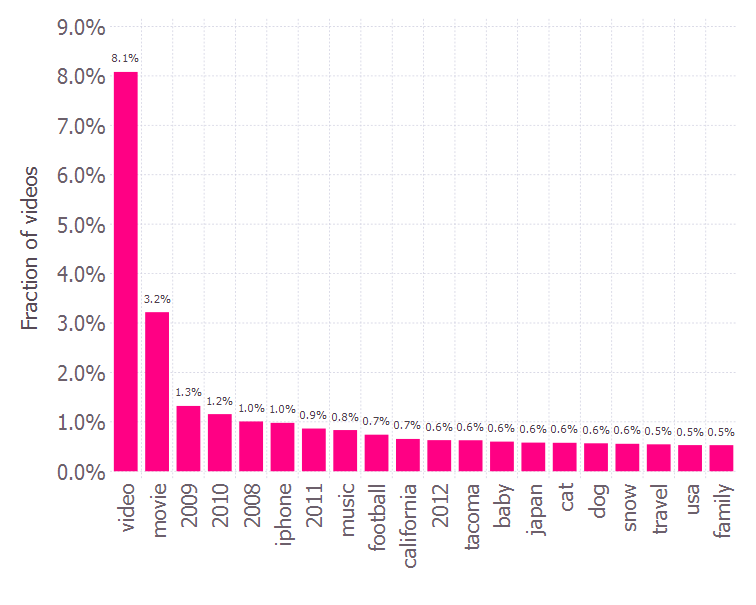} &
		\includegraphics[width=0.45\textwidth]{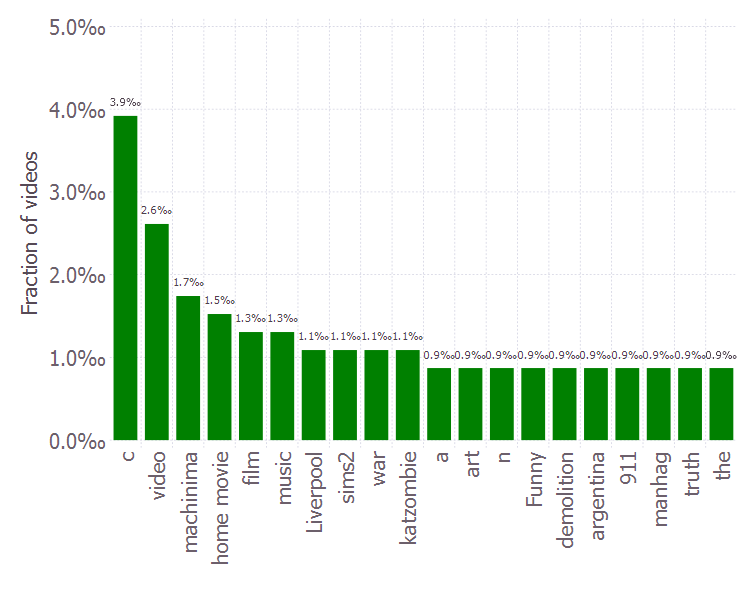}
	\end{tabular}
	\caption{Top 20 tags per collection}
	\label{fig:tags}
\end{figure}

For many applications, especially in the increasingly popular area of machine learning it is important for the videos to be annotated. Since it is impractical to show a comparison of the tags used for annotation in the different collections directly we will first content ourselves with an analysis of their number. Figure~\ref{fig:tags_per_video} shows a histogram over the number of tags per video. The IACC.3 has the fewest tags per video of all analyzed collections with the majority of its videos having no tags at all. Similarly, roughly 86\% of videos in YFCC100M have at most one tag associated with them. The two collections sourced from YouTube share a similar distribution in tag count with the data we collected but both have a lower number of videos with no tags which might be due to the process by which they were selected for inclusion in their respective collection. 
We next look at the most common tags for our data from vimeo and YouTube as well as for the IACC.3 and the YFCC100M. Since these tags are freely specifiable, meaning they are not predefined by the platform, a  de-duplication process was performed for vimeo and YouTube which caused variations in upper- and lower-case letters as well as tags preceding and trailing white space to be considered identical. For vimeo there are roughly 0.90 unique tags per video (0.73 after de-duplication)\footnote{18'097'904 total, 14'695'903 without duplicates} while the ratio for YouTube is roughly 0.986 (0.84 after de-duplication)\footnote{100'269'228 total, 85'336'249 without duplicates}. Figure~\ref{fig:tags} shows the 20 most commonly used tags for the 4 datasets as well as their relative occurrence. On vimeo, the most common tags can be grouped in to three categories: genre (`animation', `music', `art', etc.), ceremonial and religious (`wedding', `church', `jesus', etc.) and production (`canon', `after effects', `gopro'). On YouTube it is more difficult to group the most popular tags into meaningful categories. They contain genre-like labels such as `news', `tutorial' or `gameplay' but also less meaningful labels such as `tv' and `hd' as well as stop words like `the' and `of'. The only two tags present in both lists are `music' and `video'. Similarly, the YFCC100M contains many numerical tags which are in all likeliness meant to represent years. It is also interesting to note that the most common tags in the YFCC100M `video' and `movie' are about one order of magnitude more common than the most common tags in the other datasets. The tags in the IACC.3 appear to be the most diverse which is probably due to its comparatively small size. The quality of these tags is however called into question by the fact that the most common one is the letter `c'.

\clearpage

\subsection{Language}
\begin{figure}[H]
	\centering
	\begin{tabular}{cc}
		\includegraphics[width=0.5\textwidth]{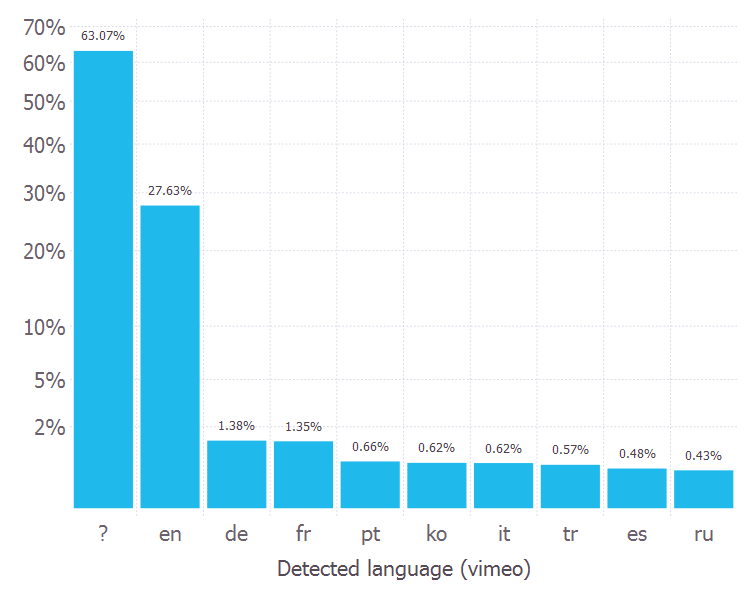} &
		\includegraphics[width=0.5\textwidth]{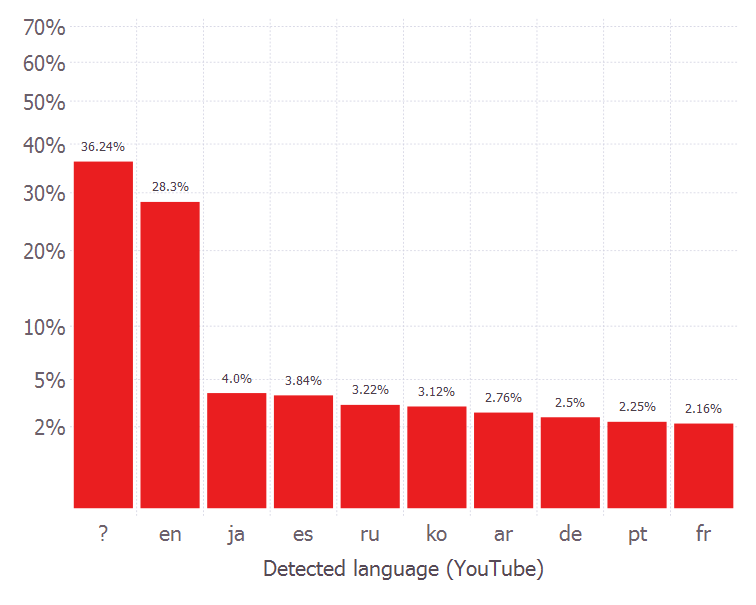} \\
		\includegraphics[width=0.5\textwidth]{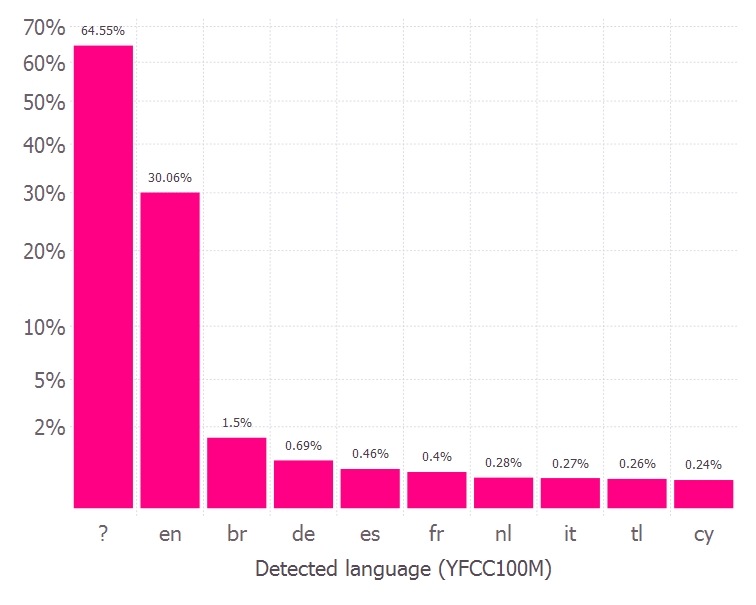} &
		\includegraphics[width=0.5\textwidth]{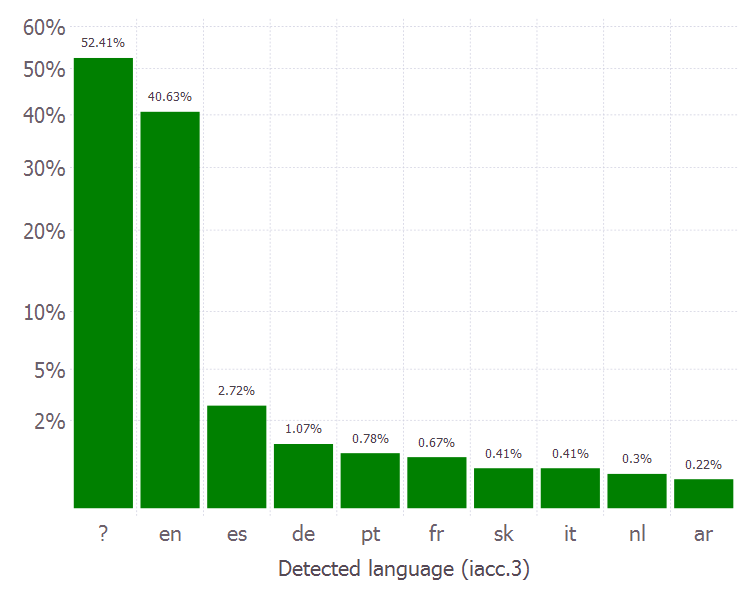}
	\end{tabular}
	\caption{Top 10 languages per platform}
	\label{fig:lang}
\end{figure}

To approximate the distribution of languages within the videos, we applied a language detection algorithm\footnote{\url{https://github.com/optimaize/language-detector}} on the title and description of the videos for the data from vimeo, YouTube and the YFCC100M. For the IACC.3, the ASR\footnote{Automatic Speaker Recognition}\cite{gauvain2010quaero,lamel2012multilingual} data provided by the authors of the collection was used. The top 10 detected languages are shown in Figure~\ref{fig:lang}. It can be seen that for the majority of videos from all datasets, language detection did not yield a result. Among the videos for which the language (or at least the language of their title and description) could be determined, English is the most common. It is followed by Japanese, Spanish and Russian on YouTube, German, French and Portuguese on vimeo, Breton, German and Spanish in the YFCC100M and Spanish, German and Portuguese in the IACC.3.

\clearpage

\section{Additional analysis of vimeo and YouTube}
\label{sec:more_analysis}
Some information we collected while gathering metadata from vimeo and YouTube is not available for the other collections we have considered in the above comparison, due to either limitations of the platform of origin or because such information would not directly be beneficial for the collection. In this section, we examine some of these properties in detail.

\subsection{Genres}
\begin{figure}[H]
	\centering
	\begin{subfigure}[b]{0.49\textwidth}
		\includegraphics[width=\textwidth]{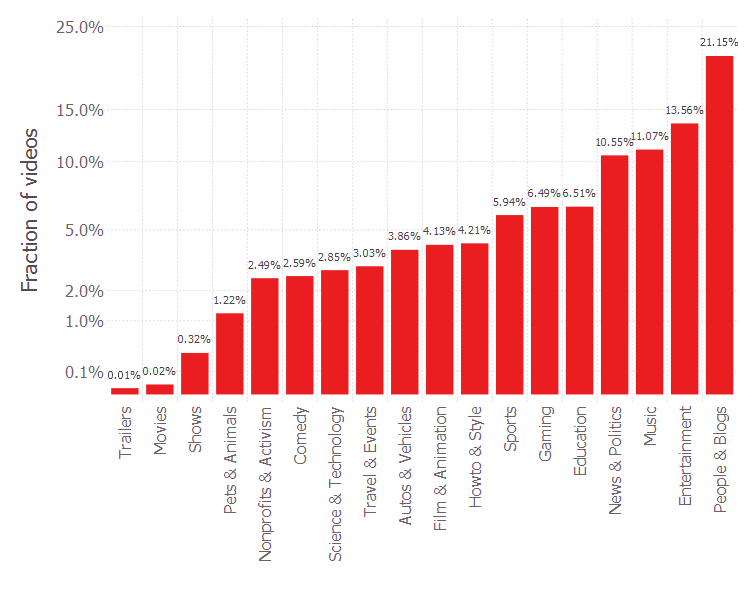}
		\caption{Genre distribution by video}
		\label{fig:genre_video}
	\end{subfigure}
	\begin{subfigure}[b]{0.49\textwidth}
		\includegraphics[width=\textwidth]{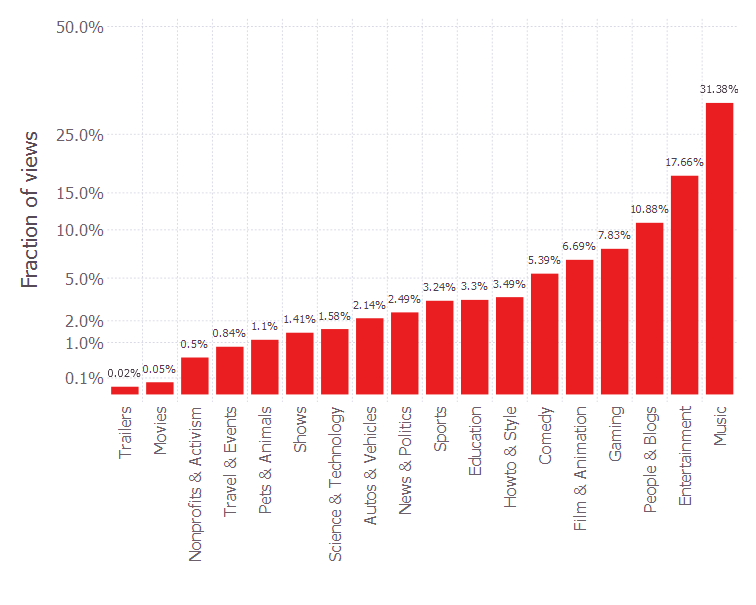}
		\caption{Genre distribution by view}
		\label{fig:genre_view}
	\end{subfigure}
	\caption{Genre distributions on YouTube}
	\label{fig:genre}
\end{figure}

When a video is uploaded to YouTube, the creator has to select to  which of the 18 predefined genres it belongs. Figure~\ref{fig:genre_video} shows the distribution of these selections. Some genres appear to attract more views than others since this distribution differs from the one depicted in Figure~\ref{fig:genre_view} which shows the fraction of views per genre. The `Music' genre, for example, is only the third most popular in terms of occurrence with roughly 11\% of videos but with over 31\% by far the most popular in terms of views. Simultaneously, while over 21\% of videos are assigned to `People \& Blogs', this genre covers less than 11\% of all views.

\subsection{Views}
\begin{figure}
	\centering
	\includegraphics[width=\textwidth]{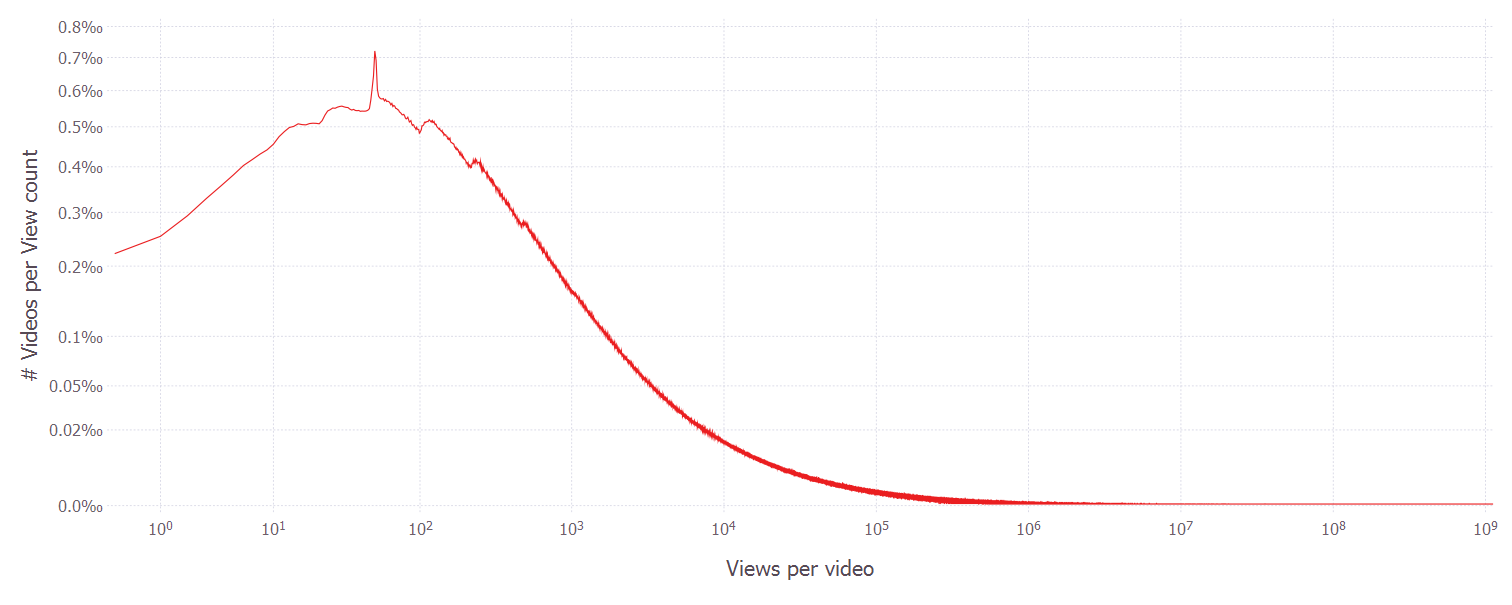}
	\caption{View distribution on YouTube}
	\label{fig:views}
\end{figure}

\begin{figure}[t]
	\centering
	\includegraphics[width=\textwidth]{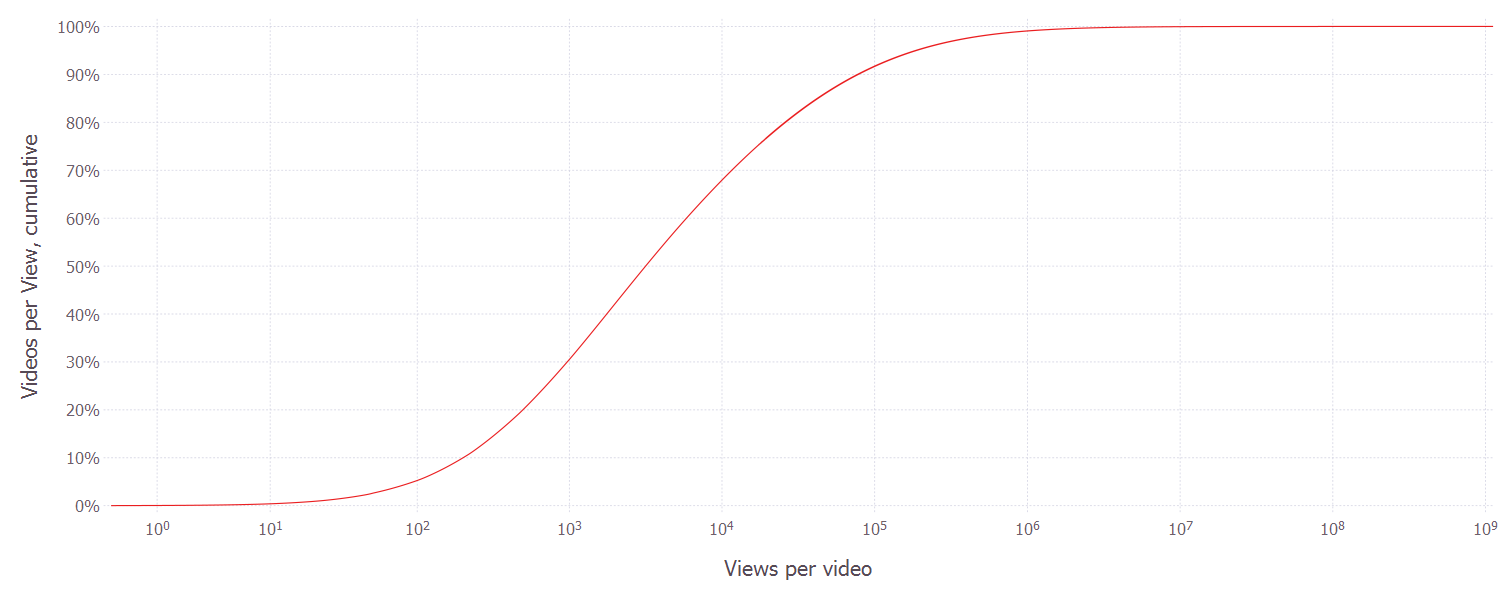}
	\caption{Cumulative view distribution on YouTube}
	\label{fig:views_cumsum}
\end{figure}

Figures~\ref{fig:views} and \ref{fig:views_cumsum} show the distribution of video views in YouTube. The most commonly occurring view count in our dataset is 50 and videos with 50 $\pm$ 2 views account for 3.85 \textperthousand{} of the dataset. 

It can be seen that many videos (almost 40\%) have between 1'000 and 10'000 views, while roughly 30\% of videos have 1'000 views or fewer. The average view count per video is 65'125 while the median lies at 3'161. This large discrepancy can be attributed to the fact that only 0.94\% of the videos have one million views or more and the 1 \textperthousand{} most viewed videos account for roughly 26.7\% of all views. Since we have videos with 0 views in the dataset (0.22 \textperthousand{}) as well as a large number of videos with only few views, we expect to have no or no significant bias towards videos with a large number of views which might have been introduced by the employed crawling strategy.

\begin{figure}[h]
	\centering
	\includegraphics[width=\textwidth]{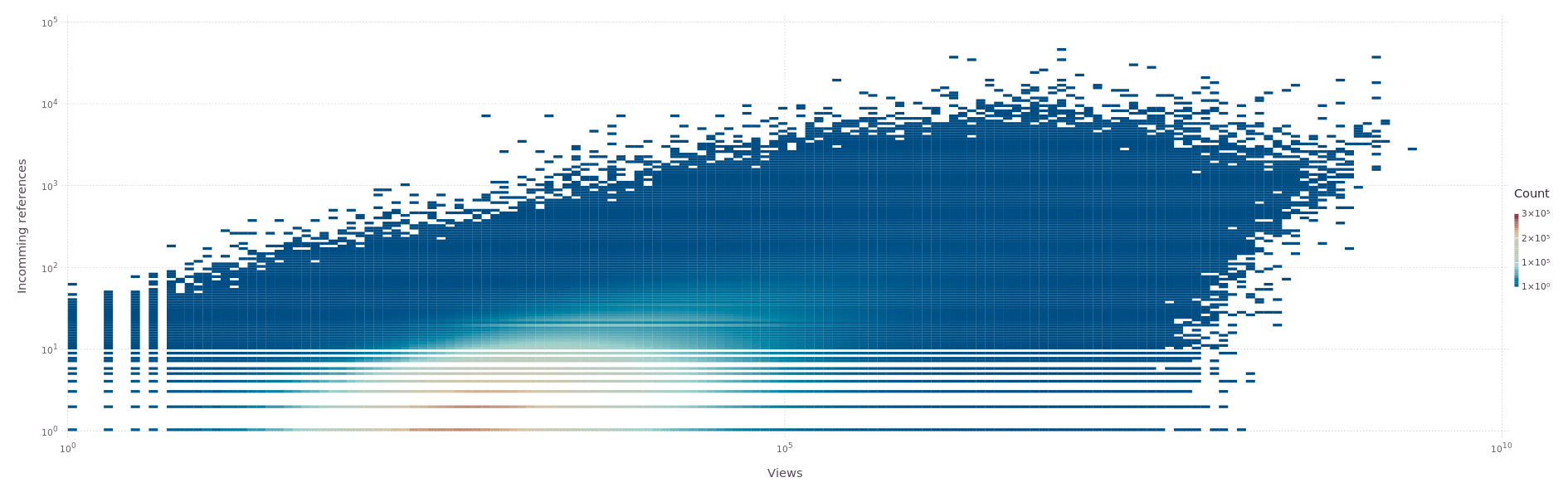}
	\caption{Distribution of views and video references on YouTube}
	\label{fig:views_references}
\end{figure}

In Figure~\ref{fig:views_references} we examine the relationship between the number of times a video is in the list of suggestions from another page and the number of its views. There appears to be little correlation between the two measures. Since this recording is a single snapshot in time, it is however difficult to argue if a higher number of views might influence the number of times a video is suggested (or, for that matter, the other way around) since a video which was popular in the past might not have been anymore at the time of our recording.

\clearpage

\subsection{Licenses}

\begin{figure}[H]
	\centering
	\begin{tabular}{cc}
		\includegraphics[width=0.5\textwidth]{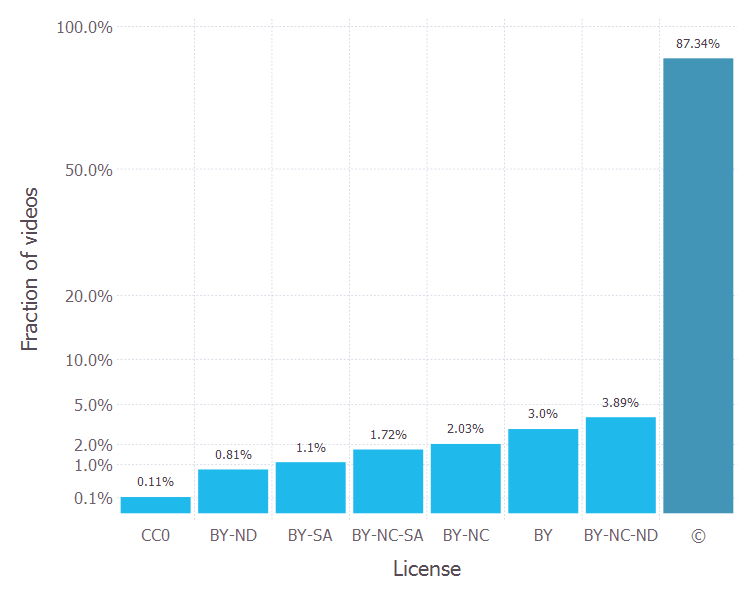} &
		\includegraphics[width=0.5\textwidth]{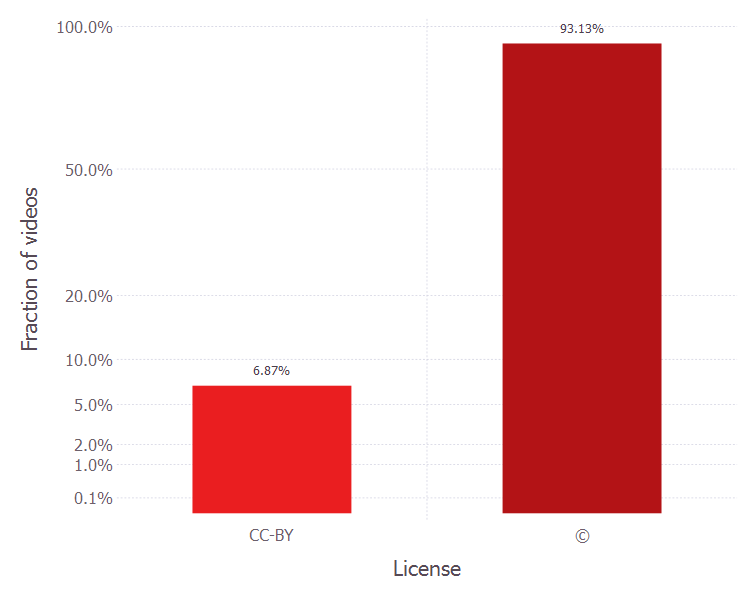}
	\end{tabular}
	\caption{Distribution of video licenses}
	\label{fig:license}
\end{figure}

Both considered platforms allow a creator some flexibility in the specification of free video licenses in addition to the default \emph{all rights reserved}. While vimeo supports all creative commons licenses, YouTube offers only the CC-BY or creative commons attribution license. Figure~\ref{fig:license} shows the distribution of these licenses on both platforms. It can be seen that vimeo has an overall higher percentage of videos with a creative commons license than YouTube. This might be due to the limited choice in licenses offered by YouTube which do not offer the possibility to label a video to be free for non-commercial use for example.

\pagebreak

\section{Discussion}
\label{sec:discussion}
When looking at the data presented in Sections~\ref{sec:analysis} and \ref{sec:more_analysis}, it becomes clear that web video has not only significantly gained in size but also in diversity. Increases can be observed along all observed axis such as resolution, duration, or upload rate. The latter is also responsible for the accelerating increase in overall content.

When comparing the different datasets, noticeable differences can be found. We find quite significant differences in the `age'-distribution, meaning the dates at which the elements of the collections have been made publicly available. This is to be expected to a certain degree as some of the collections have been created several years ago and can therefore not include elements which did not exist at that time. The source of the content, and the criteria for inclusion in the collection can account for the remaining age differences. Source-specific differences can be seen clearly in the distribution of video duration, with data from vimeo and YouTube spanning a much wider range than the videos on flickr and the elements from the IACC.3 having a significantly narrower range still. For the collections which did feature different video resolutions, we also found them to be much less diverse than what we observe on vimeo. For all analyzed datasets, English is the most commonly detected language in the video titles and descriptions. This consistency only concerns the most common language however, as from the 2$^{nd}$ most common language onward the distribution looks very different. The distribution of tags is also very different among collections, indicating large differences in the overall composition of semantic video content. This was, however, not part of our analysis.
We were also able to see that videos which has been published with a creative commons license shows differences in distribution of upload rate and duration which would also imply at least some differences in the overall content.

When focusing on the data we collected from vimeo and YouTube, it can be seen that web video not only underwent changes in the past but is still changing. Upload rates show an upward trend, so do video duration and resolution, with the aspect ratio moving towards wider screen formats. These changes together inevitably lead to an acceleration in the increase of video data over time.

Even without analyzing the content of the videos in detail, it is possible to observe differences between the three large platforms (flickr, vimeo and YouTube) which served as source for this analysis. While vimeo and YouTube both share similarities in video duration and upload pattern, flickr does not exhibit such similarities. While some of these differences might be due to the inherent differences in the possibilities offered by the platforms (vimeo for example allows for arbitrary video resolutions while YouTube does not), most common tags further strengthen the assumption that the communities which produce the content on the different platforms are rather heterogeneous and different in the way they use the corresponding platform and the content they share.

\section{Dataset}
\label{sec:dataset}
We provide the metadata collected from vimeo and YouTube via \url{http://download-dbis.dmi.unibas.ch/WWIN/} in the form of two PostgreSQL data\-base backups, one per platform. The dataset collected from YouTube contains additional data which was not used in the analysis presented in this paper, such as information on the `channels' of the video creators.

\section{Conclusions}
\label{sec:conclusion}
In this paper, we presented insights gained from the analysis of the metadata from 120 million videos from two popular web video platforms and their comparison to the properties of commonly used video collections. We showed similarities and differences between these datastes and were able to demonstrate that the analyzed video collections are not able to accurately represent the current state of web video as found `in the wild'. However, this property would be essential to use existing collections in a large variety of various applications (e.g., for evaluating the suitability of algorithms, libraries, or systems for multimedia data management, multimedia retrieval, machine learning, etc., to name just a few potential applications). 
The dataset we collected for this analysis could prove useful in future research on the increasingly relevant topic of web video and thus overcome the drawback of existing collections in terms of their representativeness, which is why we offer it to the research community for further study.

\section*{Acknowledgements}
This work was partly supported by the Chist-Era project IMOTION with contributions from the Swiss National Science Foundation (SNSF, contract no.\ 20CH21\_151571).

\clearpage

\bibliographystyle{plain}
\bibliography{biblipgraphy}

\begin{thebibliography}{10}
\expandafter\ifx\csname url\endcsname\relax
  \def\url#1{\texttt{#1}}\fi
\expandafter\ifx\csname urlprefix\endcsname\relax\def\urlprefix{URL }\fi
\providecommand{\bibinfo}[2]{#2}
\providecommand{\eprint}[2][]{\url{#2}}

\bibitem{da2016near}
\bibinfo{author}{Da~Silva, H.~B.} \emph{et~al.}
\newblock \bibinfo{title}{Near-duplicate video detection based on an
  approximate similarity self-join strategy}.
\newblock In \emph{\bibinfo{booktitle}{14th International Workshop on
  Content-based Multimedia Indexing}} (\bibinfo{year}{2016}).

\bibitem{cao2016web}
\bibinfo{author}{Cao, J.}, \bibinfo{author}{Zhang, Y.}, \bibinfo{author}{Ji,
  R.}, \bibinfo{author}{Xie, F.} \& \bibinfo{author}{Su, Y.}
\newblock \bibinfo{title}{Web video topics discovery and structuralization with
  social network}.
\newblock \emph{\bibinfo{journal}{Neurocomputing}}
  \textbf{\bibinfo{volume}{172}}, \bibinfo{pages}{53--63}
  (\bibinfo{year}{2016}).

\bibitem{li2016online}
\bibinfo{author}{Li, G.}, \bibinfo{author}{Jiang, S.}, \bibinfo{author}{Zhang,
  W.}, \bibinfo{author}{Pang, J.} \& \bibinfo{author}{Huang, Q.}
\newblock \bibinfo{title}{Online web video topic detection and tracking with
  semi-supervised learning}.
\newblock \emph{\bibinfo{journal}{Multimedia Systems}}
  \textbf{\bibinfo{volume}{22}}, \bibinfo{pages}{115--125}
  (\bibinfo{year}{2016}).

\bibitem{das2013thousand}
\bibinfo{author}{Das, P.}, \bibinfo{author}{Xu, C.}, \bibinfo{author}{Doell,
  R.~F.} \& \bibinfo{author}{Corso, J.~J.}
\newblock \bibinfo{title}{A thousand frames in just a few words: Lingual
  description of videos through latent topics and sparse object stitching}.
\newblock In \emph{\bibinfo{booktitle}{Computer Vision and Pattern Recognition
  (CVPR), 2013 IEEE Conference on}}, \bibinfo{pages}{2634--2641}
  (\bibinfo{organization}{IEEE}, \bibinfo{year}{2013}).

\bibitem{ryoo2013first}
\bibinfo{author}{Ryoo, M.~S.} \& \bibinfo{author}{Matthies, L.}
\newblock \bibinfo{title}{First-person activity recognition: What are they
  doing to me?}
\newblock In \emph{\bibinfo{booktitle}{Computer Vision and Pattern Recognition
  (CVPR), 2013 IEEE Conference on}}, \bibinfo{pages}{2730--2737}
  (\bibinfo{organization}{IEEE}, \bibinfo{year}{2013}).

\bibitem{geisler2000open}
\bibinfo{author}{Geisler, G.} \& \bibinfo{author}{Marchionini, G.}
\newblock \bibinfo{title}{The open video project: research-oriented digital
  video repository}.
\newblock In \emph{\bibinfo{booktitle}{Proceedings of the fifth ACM conference
  on Digital libraries}}, \bibinfo{pages}{258--259}
  (\bibinfo{organization}{ACM}, \bibinfo{year}{2000}).

\bibitem{OSVC1}
\bibinfo{author}{Rossetto, L.}, \bibinfo{author}{Giangreco, I.} \&
  \bibinfo{author}{Schuldt, H.}
\newblock \bibinfo{title}{{OSVC} -- {O}pen {S}hort {V}ideo {C}ollection 1.0}.
\newblock \bibinfo{type}{Tech. Rep.} \bibinfo{number}{CS-2015-002},
  \bibinfo{institution}{University of Basel} (\bibinfo{year}{2015}).

\bibitem{over2009creating}
\bibinfo{author}{Over, P.}, \bibinfo{author}{Awad, G.},
  \bibinfo{author}{Smeaton, A.~F.}, \bibinfo{author}{Foley, C.} \&
  \bibinfo{author}{Lanagan, J.}
\newblock \bibinfo{title}{Creating a web-scale video collection for research}.
\newblock In \emph{\bibinfo{booktitle}{Proceedings of the 1st workshop on
  Web-scale multimedia corpus}}, \bibinfo{pages}{25--32}
  (\bibinfo{organization}{ACM}, \bibinfo{year}{2009}).

\bibitem{song2011multiple}
\bibinfo{author}{Song, J.}, \bibinfo{author}{Yang, Y.}, \bibinfo{author}{Huang,
  Z.}, \bibinfo{author}{Shen, H.~T.} \& \bibinfo{author}{Hong, R.}
\newblock \bibinfo{title}{Multiple feature hashing for real-time large scale
  near-duplicate video retrieval}.
\newblock In \emph{\bibinfo{booktitle}{Proceedings of the 19th ACM
  international conference on Multimedia}}, \bibinfo{pages}{423--432}
  (\bibinfo{organization}{ACM}, \bibinfo{year}{2011}).

\bibitem{thomee2015yfcc100m}
\bibinfo{author}{Thomee, B.} \emph{et~al.}
\newblock \bibinfo{title}{The new data and new challenges in multimedia
  research}.
\newblock \emph{\bibinfo{journal}{arXiv preprint arXiv:1503.01817}}
  (\bibinfo{year}{2015}).

\bibitem{wu2009real}
\bibinfo{author}{Wu, X.}, \bibinfo{author}{Ngo, C.-W.},
  \bibinfo{author}{Hauptmann, A.~G.} \& \bibinfo{author}{Tan, H.-K.}
\newblock \bibinfo{title}{Real-time near-duplicate elimination for web video
  search with content and context}.
\newblock \emph{\bibinfo{journal}{IEEE Transactions on Multimedia}}
  \textbf{\bibinfo{volume}{11}}, \bibinfo{pages}{196--207}
  (\bibinfo{year}{2009}).

\bibitem{cao2009mcg}
\bibinfo{author}{Cao, J.} \emph{et~al.}
\newblock \bibinfo{title}{Mcg-webv: A benchmark dataset for web video
  analysis}.
\newblock \emph{\bibinfo{journal}{Beijing: Institute of Computing Technology}}
  \textbf{\bibinfo{volume}{10}}, \bibinfo{pages}{324--334}
  (\bibinfo{year}{2009}).

\bibitem{marszalek09}
\bibinfo{author}{Marsza{\l}ek, M.}, \bibinfo{author}{Laptev, I.} \&
  \bibinfo{author}{Schmid, C.}
\newblock \bibinfo{title}{Actions in context}.
\newblock In \emph{\bibinfo{booktitle}{IEEE Conference on Computer Vision \&
  Pattern Recognition}} (\bibinfo{year}{2009}).

\bibitem{cobarzaninteractive}
\bibinfo{author}{Cob{\^a}rzan, C.} \emph{et~al.}
\newblock \bibinfo{title}{Interactive video search tools: a detailed analysis
  of the video browser showdown 2015}.
\newblock \emph{\bibinfo{journal}{Multimedia Tools and Applications}}
  \bibinfo{pages}{1--33}.

\bibitem{algur2016web}
\bibinfo{author}{Algur, S.~P.} \& \bibinfo{author}{Bhat, P.}
\newblock \bibinfo{title}{Web video mining: Metadata predictive analysis using
  classification techniques}.
\newblock \emph{\bibinfo{journal}{International Journal of Information
  Technology and Computer Science (IJITCS)}} \textbf{\bibinfo{volume}{8}},
  \bibinfo{pages}{69} (\bibinfo{year}{2016}).

\bibitem{aggarwal2014mining}
\bibinfo{author}{Aggarwal, N.}, \bibinfo{author}{Agrawal, S.} \&
  \bibinfo{author}{Sureka, A.}
\newblock \bibinfo{title}{Mining youtube metadata for detecting privacy
  invading harassment and misdemeanor videos}.
\newblock In \emph{\bibinfo{booktitle}{Privacy, Security and Trust (PST), 2014
  Twelfth Annual International Conference on}}, \bibinfo{pages}{84--93}
  (\bibinfo{organization}{IEEE}, \bibinfo{year}{2014}).

\bibitem{imc2007cha}
\bibinfo{author}{Cha, M.}, \bibinfo{author}{Kwak, H.},
  \bibinfo{author}{Rodriguez, P.}, \bibinfo{author}{Ahn, Y.-Y.} \&
  \bibinfo{author}{Moon, S.}
\newblock \bibinfo{title}{{I Tube, You Tube, Everybody Tubes: Analyzing the
  World's Largest User Generated Content Video System}}.
\newblock In \emph{\bibinfo{booktitle}{ACM Internet Measurement Conference}}
  (\bibinfo{year}{2007}).

\bibitem{gill2007youtube}
\bibinfo{author}{Gill, P.}, \bibinfo{author}{Arlitt, M.}, \bibinfo{author}{Li,
  Z.} \& \bibinfo{author}{Mahanti, A.}
\newblock \bibinfo{title}{Youtube traffic characterization: a view from the
  edge}.
\newblock In \emph{\bibinfo{booktitle}{Proceedings of the 7th ACM SIGCOMM
  conference on Internet measurement}}, \bibinfo{pages}{15--28}
  (\bibinfo{organization}{ACM}, \bibinfo{year}{2007}).

\bibitem{zink2009characteristics}
\bibinfo{author}{Zink, M.}, \bibinfo{author}{Suh, K.}, \bibinfo{author}{Gu, Y.}
  \& \bibinfo{author}{Kurose, J.}
\newblock \bibinfo{title}{Characteristics of youtube network traffic at a
  campus network--measurements, models, and implications}.
\newblock \emph{\bibinfo{journal}{Computer networks}}
  \textbf{\bibinfo{volume}{53}}, \bibinfo{pages}{501--514}
  (\bibinfo{year}{2009}).

\bibitem{KarpathyCVPR14}
\bibinfo{author}{Karpathy, A.} \emph{et~al.}
\newblock \bibinfo{title}{Large-scale video classification with convolutional
  neural networks}.
\newblock In \emph{\bibinfo{booktitle}{CVPR}} (\bibinfo{year}{2014}).

\bibitem{choi2016analysis}
\bibinfo{author}{Choi, J.-H.} \& \bibinfo{author}{Lee, J.-S.}
\newblock \bibinfo{title}{Analysis of spatial, temporal, and content
  characteristics of videos in the yfcc100m dataset}.
\newblock In \emph{\bibinfo{booktitle}{Proceedings of the 2016 ACM Workshop on
  Multimedia COMMONS}}, \bibinfo{pages}{27--34} (\bibinfo{organization}{ACM},
  \bibinfo{year}{2016}).

\bibitem{gauvain2010quaero}
\bibinfo{author}{Gauvain, J.-L.}
\newblock \bibinfo{title}{The quaero program: Multilingual and multimedia
  technologies}.
\newblock In \emph{\bibinfo{booktitle}{International Workshop on Spoken
  Language Translation (IWSLT) 2010}} (\bibinfo{year}{2010}).

\bibitem{lamel2012multilingual}
\bibinfo{author}{Lamel, L.}
\newblock \bibinfo{title}{Multilingual speech processing activities in quaero:
  Application to multimedia search in unstructured data.}
\newblock In \emph{\bibinfo{booktitle}{Baltic HLT}}, \bibinfo{pages}{1--8}
  (\bibinfo{year}{2012}).

\end{thebibliography}

\end{document}